\documentclass[unsortedaddress,onecolumn, nofootinbib,10pt]{revtex4}
\usepackage[utf8]{inputenc}
\usepackage{amsmath,empheq,color}
\usepackage{amsfonts}
\usepackage{amsthm}
\usepackage{amssymb}
\usepackage{graphicx}
\usepackage{hyperref}
\usepackage[normalem]{ulem}
\usepackage{epigraph}
\usepackage{mathrsfs}
\usepackage{bbm}
\usepackage{wrapfig}
\usepackage{lineno}
\usepackage{scalerel,stackengine}

\begin{document}

\title{On the geometric phase for Gaussian states}

\author{Angel Garcia-Chung}
\email{alechung@xanum.uam.mx}

\affiliation{Departamento de F\'isica, Universidad Aut\'onoma Metropolitana - Iztapalapa\\
San Rafael Atlixco 186, Ciudad de M\'exico 09340, M\'exico}

\begin{abstract}
We show the explicit expression of the geometric phase for $n$-partite Gaussian states. In our analysis, the covariance matrix can be obtained as a boundary term of the geometric phase. 
\end{abstract}

\maketitle

\section{Introduction}

The covariance matrix of Gaussian states plays a prominent role detecting entanglement and squeezing in quantum states. The entanglement, which is one of the main features of quantum mechanics, not only due its philosophical implications but also for its applications in many different areas \cite{bennett2000quantum, divincenzo1995quantum, bouwmeester1997experimental,  pan2001entanglement,  axline2018demand, adesso2007entanglement, horodecki2009quantum, richens2017entanglement}, can be detected using several criteria \cite{peres1996separability, horodecki1997separability, simon2000peres, duan2000inseparability, werner2001bound, giedke2001separability}. In the case of continuous variable (CV) systems, these criteria rely on some features of the covariance matrix \cite{simon2000peres, duan2000inseparability, werner2001bound, giedke2001separability}.



On the other hand, squeezed states can be used to improve the sensitivity of measurement devices beyond the usual quantum noise limits \cite{simon1988gaussian, braunstein2005quantum, walls2007quantum, adesso2014continuous, ma1990multimode, schnabel2017squeezed, pirandola2009correlation, weedbrook2012gaussian}. These states are a particular type of Gaussian states, i.e., those states whose Wigner functions are Gaussian functions on the phase space \cite{simon1988gaussian, braunstein2005quantum, walls2007quantum, adesso2014continuous}. Due to the covariance matrix contains all the information related with the uncertainty relations, it is used to determine whether a state is squeezed or not \cite{simon1988gaussian}.

 The squeezed states are generated by the squeeze operators which can be considered as elements of the unitary representation of the symplectic group $Sp(2n, \mathbb{R})$, specifically, those close to the group identity \cite{moshinsky1971linear, Arvind:1995ab, torre2005linear, wolf2016development, de2011symplectic}. When the squeeze operator acts on the vacuum state the resulting squeezed state is named vacuum squeezed state. A distinctive feature of the vacuum squeezed state is the expression for its covariance matrix
\begin{equation}
{\bf V}^{(2)} = \frac{1}{2} {\bf M} \; {\bf M}^T, \label{CVIntro}
\end{equation}
\noindent where the matrix ${\bf M}$ is an element of the symplectic group $Sp(2n, \mathbb{R})$ \cite{pirandola2009correlation, adesso2014continuous}. We can notice in (\ref{CVIntro}) the connection between the covariance matrix and the symplectic group $Sp(2n, \mathbb{R})$. This connection paves the way to the following question: is there a direct relation between the covariance matrix and the geometric phase for vacuum squeezed states?  Let us provide some context for this question.

Geometric phases are obtained when a physical system evolve through a set of states \cite{hariharan2005geometric, chruscinski2012geometric}. These phases are ubiquitous in many different research areas \cite{hariharan2005geometric, sachdev2007quantum, cabra2012modern, chruscinski2012geometric}, particularly, in quantum optics where they have been studied extensively at both levels, theoretically and experimentally \cite{hariharan2005geometric, chaturvedi1987berry, simon1993bargmann, kwiat1991observation, mukunda1993quantum, brendel1995geometric, strekalov1997two, seshadri1997geometric, mendas1997pancharatnam, fuentes2000proposal, de2002squeezing}. Among the manifestations of geometric phases in optics, we are interested in the phase resulting from a cycle of changes in squeezed states of light \cite{hariharan2005geometric, chaturvedi1987berry, Chiao, seshadri1997geometric}. For simplicity, in this work we will refer to this phase as: {\it squeeze phase}.

For a system with a single photon ($n=1$), the squeeze phase results from a cyclic variation of the parameter $\theta \in [0, 2 \pi]$. Recall that $\theta$ together with $r$, define the complex parameter $\zeta = r e^{i \theta}$ which in turn, define the squeeze operator $\widehat{S}(\zeta)$ \cite{walls2007quantum, adesso2014continuous}. In the single photon scenario and for the squeezed state in the $j$-th energy level, the geometric phase $\gamma^{(1)}_j$, was reported in \cite{chaturvedi1987berry}. In this notation, the index ${}^{(1)}$ for the phase, labels the number of photons. In the case of general coherent squeezed states, the geometric phase $\gamma^{(n=1)}_j$, and the phase difference, $\Delta \gamma^{(n=1)}_j = \gamma^{(n=1)}_j -  \gamma^{(n=1)}_{j-1}$, was treated in \cite{seshadri1997geometric}. As can be seen and to the best of the author's knowledge, only the single photon cases have been considered. Hence, the analysis of the squeeze phase for a multi-photon system, i.e., $\gamma^{(n>1)}_0$, is still open. This becomes more relevant if we consider that the experimental results \cite{brendel1995geometric, hariharan2005geometric} imply that the squeeze phase for pairs of photons with parallel polarizations acquire twice the geometric phase of single photons.

On the other hand, due to the squeeze operator can be seen as an element of the unitary representation of $Sp(2n,\mathbb{R})$, a cyclic variation of $\theta$ yields a cyclic variation of ${\bf M}$. As a result, due to the connection in (\ref{CVIntro}) between ${\bf V}^{(2)}$ and ${\bf M}$, we might expect also a cyclic variation of ${\bf V}^{(2)}$. Moreover, we can go beyond and generalize our analysis and consider an arbitrary Gaussian state. That being said, we consider that the motivation for our previous question seems now more natural. 

The purpose of this work is to determine the geometric phase of a $n$-partite Gaussian state and for an arbitrary symplectic matrix ${\bf M}$. Our results will be valid for both, systems described with quadrature operators and for the standard dimension-full operators $\widehat{x}$ and $\widehat{p}$ of the harmonic oscillator. We will show that the covariance matrix ${\bf V}^{(2)}$ for a general Gaussian state can be seen as a boundary term. To do so, we briefly describe in Section (\ref{preliminaries}) the mathematical tools regarding the symplectic group analysis and its group action in the phase space $\mathbb{R}^{2n}$. Then, in Section (\ref{CovarianceMatrix}) we summarize the main aspects related with its unitary representation. We also derive in this section the explicit form of the covariance matrix in terms of the symplectic group elements. In Section (\ref{BerryPhase}) we calculate the explicit relation between the covariance matrix and the geometric phase and in Section (\ref{Discussion}) we provide our conclusions.

\section{Preliminaries} \label{preliminaries}
The symplectic group $Sp(2n, \mathbb{R})$ is given by $2n \times 2n$ matrices ${\bf M}$ satisfying the equation
\begin{equation}
{\bf \Omega} = {\bf M} \; {\bf \Omega} \; {\bf M}^T, \label{SpCond}
\end{equation}
\noindent where ${\bf M}^T$ is the transpose matrix of ${\bf M}$ and ${\bf \Omega}$ is defined as
\begin{equation}
{\bf \Omega} = \left( \begin{array}{cc} {\bf 0} & {\bf 1} \\ - {\bf 1} & {\bf 0} \end{array}\right).
\end{equation}
\noindent  Here ${\bf 0}$ and ${\bf 1}$ are the $n\times n$ zero and identity matrix respectively. Matrix ${\bf M}$ can be written in block form as follows
\begin{equation}
{\bf M} = \left( \begin{array}{cc} {\bf A} & {\bf B} \\ {\bf C} & {\bf D} \end{array}\right),  \label{NoTildeMatrix}
\end{equation}
\noindent where ${\bf A}$, ${\bf B}$, ${\bf C}$ and ${\bf D}$ are $n \times n$ real matrices. The condition (\ref{SpCond}) now reads as
\begin{equation}
{\bf A} {\bf D}^T - {\bf B} {\bf C}^T = {\bf 1}, \quad {\bf A} {\bf B}^T = {\bf B} {\bf A}^T, \quad {\bf C}{ \bf D}^T = {\bf D} {\bf C}^T. \label{Sp1Coord}
\end{equation}

Additionally, the group elements close to the identity can also be related with the elements of the Lie algebra $sp(2n, \mathbb{R})$ via the exponential map
\begin{equation}
{\bf M}({\bf L})= e^{ {\bf \Omega} {\bf L}}. \label{ExpMap}
\end{equation}
\noindent where ${\bf L} \in sp(2n, \mathbb{R})$ is a real symmetric $2n \times 2n$ matrix. Here, ${\bf M}({\bf L})$  refers to the explicit the relation between the group element ${\bf M}$ and its associated Lie algebra matrix ${\bf L}$ via the exponential map. There are, of course, elements in $Sp(2n, \mathbb{R})$ that are not close to the identity and therefore, cannot be written as the exponential map of any symmetric matrix ${\bf L}$. In the appendix \ref{LAAnalysis} we derive the relation ${\bf M}(\bf L)$ between the group $Sp(4,\mathbb{R})$ and its Lie algebra $sp(4, \mathbb{R})$. This result will be very useful in the analysis of the symplectic group representation for bi-partite systems.

The symplectic group is used to provide the action of the linear canonical transformations on the phase space $(\mathbb{R}^{2n}, \{, \})$ where $\{, \}$ is the standard Poisson bracket. The group action on $\mathbb{R}^{2n}$ is then given as
\begin{equation}
\vec{\bf X}'^T = {\bf M} \, \vec{\bf X}^T, \label{1Coord}
\end{equation}
\noindent where $\vec{\bf X}^T = (\vec{q} \; \vec{p})^T$ and $\vec{q}=(q_1, q_2, \dots, q_n)$ and $\vec{p}=(p_1, p_2, \dots, p_n)$ are the coordinates on the space $\mathbb{R}^{2n}$. Using these coordinates, the Poisson bracket can be written as
\begin{equation}
\{  \vec{\bf X}^T , \vec{\bf X} \} = {\bf \Omega}. \label{PoissonB}
\end{equation}
\noindent If we now impose that (\ref{PoissonB}) also holds for ${\vec{\bf X}}'$ then the relation in (\ref{1Coord}) yields the condition (\ref{SpCond}). 

This way of defining the symplectic group action on the phase space is very useful to obtain a unitary representation on a Hilbert space. However, to study entanglement conditions a more convenient form of the group action on the phase space $\mathbb{R}^{2n}$ is required. Consider the different array $\vec{Y}^T = ( q_1, \, p_1 , \, q_2 , \, p_2 , \, \hdots , \, q_n , \, p_n )$. Then, similarly to (\ref{1Coord}) we consider the group action to be of the form
\begin{equation}
\vec{\bf Y}'^T = \widetilde{\bf M}\; \vec{\bf Y}^T, \label{2Coord}
\end{equation}
\noindent where $\widetilde{\bf M}$ is the new form of the symplectic matrix. We now insert the expression (\ref{2Coord}), {\it mutatis mutandis}, in the Poisson bracket (\ref{PoissonB}) to obtain the following symplectic group condition
\begin{equation}
\left( \begin{array}{cccc} {\bf J} & {\bf 0} & \dots & {\bf 0} \\ {\bf 0} & {\bf J} & \dots & {\bf 0} \\ \vdots  & \vdots & \ddots & \vdots \\ {\bf 0} & {\bf 0} & \dots & {\bf J} \end{array} \right) = \widetilde{\bf M} \left( \begin{array}{cccc} {\bf J} & {\bf 0} & \dots & {\bf 0} \\ {\bf 0} & {\bf J} & \dots & {\bf 0} \\ \vdots  & \vdots & \ddots & \vdots \\ {\bf 0} & {\bf 0} & \dots & {\bf J} \end{array} \right)  \widetilde{\bf M} ^T, \label{2SpCond}
\end{equation}
\noindent where ${\bf J}$ is the $2\times 2$ matrix given by ${\bf J} = \left( \begin{array}{cc} 0 & 1 \\ -1 & 0 \end{array}\right)$. Analogously to the matrix ${\bf M}$, the matrix $\widetilde{\bf M}$ can also be written in block form as
\begin{equation}
\widetilde{\bf M} = \left( \begin{array}{cccc} {\bf A}_{11} & {\bf A}_{12} & \dots & {\bf A}_{1n} \\ {\bf A}_{21} & {\bf A}_{22} & \dots & {\bf A}_{2n} \\ \vdots & \vdots & \ddots & \vdots \\ {\bf A}_{n1} & {\bf A}_{n2} & \dots & {\bf A}_{nn} \end{array}\right) , \label{TildeMatrix}
\end{equation}
\noindent where ${\bf A}_{i j}$, with $i,j = 1,2, \dots, n$, are $n^2$ real $2\times 2$ matrices. The condition (\ref{2SpCond}) now reads on these block matrices as
\begin{equation}
{\bf J} = \sum^{j=n}_{j=1} {\bf A}_{i\, j} {\bf J} {\bf A}^T_{i\,j}, \quad  {\bf 0} = \sum^{k=n}_{k=1} {\bf A}_{i\, k} {\bf J} {\bf A}^T_{j\, k}, \label{Sp2Coord}
\end{equation}
\noindent for all $i \neq j$ in the second condition. 

Expressions (\ref{Sp1Coord}) and (\ref{Sp2Coord}) define different conditions for the symplectic matrices although, they describe the same Lie group \cite{adesso2014continuous}. Notice in (\ref{TildeMatrix}) that when the off-diagonal block matrices ${\bf A}_{i \neq j} = {\bf 0}$ the symplectic matrix becomes block-diagonal, i.e., 
\begin{equation}
\widetilde{\bf M}= \mbox{diag}({\bf A}_{11}, {\bf A}_{22} , \dots, {\bf A}_{nn}), \label{Entang}
\end{equation}
\noindent then, notably, as a result of this form of the group action (\ref{Sp2Coord}), each matrix ${\bf A}_{ii} $ is an element of the symplectic group $ Sp(2,\mathbb{R})$.

Both group actions (\ref{NoTildeMatrix}) and (\ref{TildeMatrix}) are related via the transformation ${\bf \Gamma}$ \cite{adesso2014continuous} which is defined as
\begin{equation}
{\bf M} = {\bf \Gamma} \; \widetilde{\bf M} \; {\bf \Gamma}^{-1}, \label{RelationbetweenMs}
\end{equation}
\noindent where ${\bf \Gamma}$ is given by
\begin{equation}
\vec{\bf X}^T = {\bf \Gamma} \, \vec{\bf Y}^T , \label{CTransformation}
\end{equation}
\noindent and is such that ${\bf \Gamma}^T  = {\bf \Gamma}^{-1}$. For example, in the case of $n=2$ the matrix transformation, denoted by $\Gamma^{(2)}$, is
\begin{equation}
\Gamma^{(2)} = \left( \begin{array}{cccc} 1 & 0 & 0 &  0 \\ 0 & 0 & 1 & 0 \\ 0 & 1 & 0 & 0 \\ 0 & 0 & 0 & 1  \end{array}\right). \label{Gamma2}
\end{equation}

Let us now move to the next section, where we briefly show the analysis to calculate the covariance matrix using the unitary representation of the symplectic group. These results will be use in the calculation of the Berry phase in (\ref{BerryPhase}). 

\section{Covariance matrix} \label{CovarianceMatrix}

As we mentioned in the introduction, the calculation of the Berry phase requires the expression of some of the covariance matrix components. For completeness, let us proceed in this section to derive the general expression for the covariance matrix of Gaussian states.

The symplectic group is a non-compact group which requires an infinite Hilbert space for its unitary representation. Let us consider the Hilbert space ${\cal H} = L^2(\mathbb{R}^n, d\vec{x})$. The unitary representation of $Sp(2n,\mathbb{R})$ is the map ${\bf M} \mapsto \widehat{C}_{\bf M}$, where $\widehat{C}_{\bf M}$ is a unitary operator $\widehat{C}_{\bf M} \in {\cal L}(\cal H)$. Here, ${\cal L}(\cal H)$ is the space of linear operators over the Hilbert space ${\cal H}$. The operator $\widehat{C}_{\bf M}$ is related with the fundamental operators $\widehat{x}_j$ and $\widehat{p}_j$ via the following condition
\begin{equation}
\widehat{C}_{\bf M} \left( \begin{array}{c} \vec{\widehat{x}} \\ \vec{\widehat{p}}\end{array} \right) \widehat{C}^{-1}_{\bf M} = {\bf M}^{-1} \left( \begin{array}{c} \vec{\widehat{x}} \\ \vec{\widehat{p}}\end{array} \right) . \label{QASG}
\end{equation}
\noindent This condition enables us to consider $\widehat{C}_{\bf M}$ as the (unitary) representation of $Sp(2n, \mathbb{R})$ on ${\cal H}$ \cite{moshinsky1971linear,  torre2005linear, wolf2016development, de2011symplectic}. The action of this operator on the Hilbert space is given by
\begin{equation}
\widehat{C}_{\bf M} \Psi(\vec{x}) = \int d\vec{x}' C_{\bf M}(\vec{x} , \vec{x}') \Psi(\vec{x}') =: \Psi_{\bf M}(\vec{x}), \label{UnitaryRep}
\end{equation}
\noindent for $\Psi(\vec{x}) \in {\cal H}$ and where the kernel of this integral operator is 
\begin{equation}
C_{\bf M}(\vec{x}, \vec{x}') = \frac{e^{ \frac{i}{2 \hbar} \left[ \vec{x}^T {\bf D} {\bf B}^{-1} \vec{x} - 2 \vec{x}'^T {\bf B}^{-1} \vec{x} + \vec{x}'^T {\bf B}^{-1} {\bf A} \vec{x}'\right]}}{\sqrt{ (2 \pi i \hbar)^n \det {\bf B}}} . \label{Kernel}
\end{equation}


It can be checked that this representation (\ref{UnitaryRep})-(\ref{Kernel}) is unitary and also, the factor $\det {\bf B}$ in (\ref{Kernel}) gives rise to a well define action even in the case where the matrix ${\bf B}$ is singular \cite{moshinsky1971linear, wolf2016development}. Moreover, the representation is valid for the entire symplectic group and not just for those elements close to the group identity. Due to we are interested in group elements where ${\bf B}$ is non-singular we omit this analysis and provide in appendix \ref{SingularCase} the analysis of the Berry phase for the very special case where ${\bf B}={\bf 0}$.

Let us proceed with the calculation of the covariance matrix. Consider the state 
\begin{equation}
| \Psi_{\bf M} \rangle = \widehat{C}_{\bf M} | 0 \rangle = \int d\vec{x} \; \Psi_{\bf M}(\vec{x}) \, | \vec{x} \rangle, \label{InitialS}
\end{equation}
\noindent where $| 0 \rangle = \int d\vec{x} \; \Psi_0(\vec{x}) \, | \vec{x} \rangle $ is the state $|0\rangle = |0\rangle_1 \otimes | 0\rangle_2 \dots | 0 \rangle_n$.  The ket $| 0 \rangle_j$ is the vacuum state of the j-th quantum harmonic oscillator. These states are Gaussian states due to the kernel (\ref{Kernel}) and moreover, they are eigenstates of the Hamiltonian
\begin{equation}
\widehat{H}_{\bf M} = \widehat{C}_{\bf M} \widehat{H} \widehat{C}^{-1}_{\bf M},
\end{equation}
\noindent where $\widehat{H} = \sum^n_j \widehat{H}_j$ is the Hamiltonian of $n-$decoupled quantum harmonic oscillators each of them characterized by the Hamiltonian $\widehat{H}_j  = \frac{1}{2 m_j} \widehat{p}^2_j + \frac{m_j \omega^2_j}{2} \widehat{q}^2_j$. 

To calculate the covariance matrix we first calculate the following amplitude
\begin{equation}
\langle \Psi_{\bf M} | \widehat{W}(\vec{a}, \vec{b}) | \Psi_{\bf M} \rangle = \langle \Psi_0 | \widehat{C}^\dagger_{\bf M} \widehat{W}(\vec{a}, \vec{b}) \widehat{C}_{\bf M} | \Psi_0 \rangle, \label{Amplitude}
\end{equation}
\noindent where $\widehat{W}(\vec{a}, \vec{b})$ is a Weyl-algebra generator whose representation on ${\cal H}$ is
\begin{equation}
\widehat{W}(\vec{a}, \vec{b}) \Psi(\vec{x}) = e^{\frac{i}{2 \hbar} \vec{a}^T \vec{b}} e^{\frac{i}{\hbar} \vec{a}^T \vec{x}} \Psi(\vec{x} + \vec{b}).
\end{equation}

A tedious but standard calculation gives the amplitude in  (\ref{Amplitude}) the following form
\begin{equation}
\langle \Psi_{\bf M} | \widehat{W}(\vec{a}, \vec{b})  | \Psi_{\bf M} \rangle  = \exp\left\{ - \frac{1}{4}   \left( \begin{array}{cc} \vec{a} & \vec{b} \end{array} \right)^T {\bf \Lambda} \left( \begin{array}{c} \vec{a} \\ \vec{b} \end{array}\right) \right\}, \label{Amplitude2}
\end{equation}
\noindent where the matrix ${\bf \Lambda}$ is given by
\begin{equation}
{\bf \Lambda} := {\bf M} \left( \begin{array}{cc} \frac{1}{\hbar^2} {\bf L}^2 & {\bf 0} \\ {\bf 0} & {\bf L}^{-2} \end{array}\right) {\bf M}^T, 
\end{equation}
\noindent and $ {\bf L} = \mbox{diag}(  l_1 , l_2 , \dots ,  l_n )$. Here, $l_j = \sqrt{\frac{\hbar}{m_j \, \omega_j}}$, with $j=1,2, \dots, n$, is the characteristic length of the quantum harmonic oscillator labelled by $j$.

The covariance matrix ${\bf V}^{(2)}$, whose components can then be written as
\begin{equation}
{\bf V}^{(2)} = \left( \begin{array}{cc} \langle \Psi_{\bf M} | \widehat{x}_j \; \widehat{x}_k  | \Psi_{\bf M} \rangle & \frac{1}{2} \langle \Psi_{\bf M} | \{ \widehat{x}_j , \widehat{p}_k \} | \Psi_{\bf M} \rangle \\  \frac{1}{2} \langle \Psi_{\bf M} | \{ \widehat{p}_j , \widehat{x}_k \} | \Psi_{\bf M} \rangle  & \langle \Psi_{\bf M} | \widehat{p}_j \; \widehat{p}_k   | \Psi_{\bf M} \rangle \end{array}\right),  \label{CVMatrixDef}
\end{equation}
\noindent can be obtained from (\ref{Amplitude2}) using the following relations
\begin{eqnarray}
\langle \Psi_{\bf M} | \widehat{x}_j \; \widehat{x}_k  | \Psi_{\bf M} \rangle = \frac{\hbar^2}{i^2} \partial^2_{ a_j a_k} \langle \Psi_{\bf M} | \widehat{W}(\vec{a}, \vec{b})  | \Psi_{\bf M} \rangle \vert_{\vec{a}, \vec{b}=0}, \label{EqCV1}\\
\frac{1}{2} \langle \Psi_{\bf M} | \{ \widehat{x}_j , \widehat{p}_k \} | \Psi_{\bf M} \rangle = \frac{\hbar^2}{i^2}  \partial^2_{a_j b_k} \langle \Psi_{\bf M} | \widehat{W}(\vec{a}, \vec{b})  | \Psi_{\bf M} \rangle \vert_{\vec{a}, \vec{b}=0},\label{EqCV2}  \\
\frac{1}{2}\langle \Psi_{\bf M} | \{ \widehat{p}_j , \widehat{x}_k \} | \Psi_{\bf M} \rangle = \frac{\hbar^2}{i^2}  \partial^2_{b_j a_k} \langle \Psi_{\bf M} |  \widehat{W}(\vec{a}, \vec{b})  | \Psi_{\bf M} \rangle \vert_{\vec{a}, \vec{b}=0}, \label{EqCV3} \\
\langle \Psi_{\bf M} |  \widehat{p}_j \; \widehat{p}_k  | \Psi_{\bf M} \rangle = \frac{\hbar^2}{i^2} \partial^2_{b_j b_k} \langle \Psi_{\bf M} |  \widehat{W}(\vec{a}, \vec{b})  | \Psi_{\bf M} \rangle \vert_{\vec{a}, \vec{b}=0}, \label{EqCV4}
\end{eqnarray}
\noindent an a straightforward calculation yields the following expression for ${\bf V}^{(2)} $ 
\begin{equation}
{\bf V}^{(2)}  = \frac{\hbar^2}{2}{\bf M} \left( \begin{array}{cc} \frac{{\bf L}^2}{\hbar^2} & {\bf 0} \\ {\bf 0} & {\bf L}^{-2} \end{array}\right) {\bf M}^T . \label{CVMatrix}
\end{equation}
\noindent It is worth to mention that the matrix for the first order moments is zero due to the symmetry of the Gaussian function in (\ref{Amplitude2}). 

In quantum optics we usually work with quadrature operators, which are dimensionless operators, and hence, can be derived by taking dimensionless parameters $l_j = 1$ and $\hbar=1$. As a result, the covariance matrix takes the simplified form
\begin{equation}
{\bf V}^{(2)}_q = \frac{1}{2} {\bf M} {\bf M}^T, \label{QCVMatrix}
\end{equation}
\noindent where the sub-index $q$ in ${\bf V}^{(2)}_q$ means we are considering quadrature operators instead of  ${\bf V}^{(2)}$ which stands for dimension-full parameters. Notice that ${\bf V}^{(2)}$ and ${\bf V}^{(2)}_q$ cannot be considered as symplectic matrices due to the $1/2$ factor. Additionally, it is worth to recall that this result is valid for systems with n-degrees of freedom and Gaussian states of the form given in (\ref{InitialS}). In the case of $n=2$, the appendix \ref{LAAnalysis} provides the general expression for the covariance matrices ${\bf V}^{(2)}$ and ${\bf V}^{(2)}_q$ using symplectic matrices ${\bf M}({\bf L})$.

We are now ready to derive, in the next section, the Berry phase for general Gaussian states of the form (\ref{InitialS}).

\section{Geometric phase for Gaussian states} \label{BerryPhase}

To derive the geometric phase let us first consider the expression for the kernel $C_{\bf M}(\vec{x}, \vec{x'})$ and notice that it does not depend on the block matrix ${\bf C}$  defined in (\ref{NoTildeMatrix}). Therefore, every variation will involve only those parameters in the kernel defined in (\ref{Kernel}). For the very special case when ${\bf B} = {\bf 0}$, the representation is rather different to (\ref{Kernel}). Due to we are interested in non-singular group elements and in order to simplify our presentation, we showed in the appendix (\ref{SingularCase}) the analysis of the geometric phase when ${\bf B} = {\bf 0}$. 

Having said this, let us consider a variation in the parameters $A_{a b}$, $B_{a b}$ and $D_{a b}$, which are the components of the block matrices ${\bf A}$, ${\bf B}$ and ${\bf D}$ respectively and recall that, the parameters $m_j$ and $\omega_j$, defined in the Hamiltonians $\widehat{H}_j$, remain fixed. With all these considerations the geometric phase can be written as 
\begin{equation}
\gamma_{\bf M} = i \int_{\cal C} \left[  \langle 0 | \widehat{C}^\dagger_{\bf M} \frac{\partial \widehat{C}_{\bf M}}{\partial A_{a b}}  | 0 \rangle \, dA_{a b} + \langle 0 | \widehat{C}^\dagger_{\bf M} \frac{\partial \widehat{C}_{\bf M}}{\partial B_{ ab}}  | 0 \rangle \, dB_{a b} + \langle 0 | \widehat{C}^\dagger_{\bf M} \frac{\partial \widehat{C}_{\bf M}}{\partial D_{ij}}  | 0 \rangle \, dD_{a b} \right], \label{BerryDef}
\end{equation}
\noindent where each of the amplitudes in (\ref{BerryDef}) have to be determined separately.

Another tedious but standard calculation yields the following expressions for these amplitudes
\begin{eqnarray} 
\langle 0 | \widehat{C}^\dagger_{\bf M} \frac{\partial \widehat{C}_{\bf M}}{\partial A_{ab}} | 0 \rangle  &=&\frac{i}{2 \hbar} \langle 0 | \widehat{x}_b \; \widehat{x}_j | 0 \rangle B^{-1}_{j a}=  \frac{i}{4 \hbar} l^2_b B^{-1}_{ba}, \label{Amp1}\\  
\langle 0 | \widehat{C}^\dagger_{\bf M} \frac{\partial \widehat{C}_{\bf M}}{\partial B_{ab}} | 0 \rangle &=&\frac{1}{2 i \hbar} B^{-1}_{b l} \langle 0 | \widehat{C}^\dagger_{\bf M} \, \widehat{x} _l \; \widehat{x}_j \widehat{C}_{\bf M} | 0  \rangle D_{j k} B^{-1}_{k a} + \frac{i}{ \hbar} B^{-1}_{b l} \langle 0 | \widehat{C}^\dagger_{\bf M} \, \widehat{x} _l \; \widehat{C}_{\bf M} \; \widehat{x}_k | 0  \rangle B^{-1}_{k a} + \frac{1}{2 i \hbar} B^{-1}_{b l} A_{l j} \langle 0 | \widehat{x}_j \; \widehat{x}_k | 0 \rangle B^{-1}_{k a} + \nonumber \\
&& - \frac{1}{2} B^{-1}_{b a} = - \frac{i}{4 \hbar} B^{-1}_{b c} A_{c d} l^2_d C_{a d} - \frac{i \hbar}{4} l^{-2}_b D_{ab},  \label{Amp2} \\
\langle 0 | \widehat{C}^\dagger_{\bf M} \frac{\partial \widehat{C}_{\bf M}}{\partial D_{ab}} | 0 \rangle &=& \frac{i}{2 \hbar} B^{-1}_{b j} \langle 0 | \widehat{C}^\dagger_{\bf M} \; \widehat{x}_j \; \widehat{x}_a \; \widehat{C}_{\bf M} | 0 \rangle = \frac{i}{4 \hbar} B^{-1}_{b c} A_{c d} l^2_d A_{a d} + \frac{i \hbar}{4} l^{-2}_b B_{a b}. \label{Amp3}
\end{eqnarray}
\noindent Here, the covariance matrix expression (\ref{CVMatrix}) is needed. The amplitudes $ \langle 0 | \widehat{x}_b \; \widehat{x}_j | 0 \rangle$ and $ \langle 0 | \widehat{C}^\dagger_{\bf M} \; \widehat{x}_b \; \widehat{x}_j \; \widehat{C}_{\bf M} | 0 \rangle$ are replaced by the expressions in (\ref{EqCV1}) and (\ref{CVMatrix}) of the covariance matrix components. For the amplitude $\langle 0 | \widehat{C}^\dagger_{\bf M} \, \widehat{x} _l \; \widehat{C}_{\bf M} \; \widehat{x}_k | 0  \rangle$ we used the expression (\ref{QASG}) \begin{equation}
\langle 0 | \widehat{C}^\dagger_{\bf M} \, \widehat{x} _l \; \widehat{C}_{\bf M} \; \widehat{x}_k | 0  \rangle = \langle 0 | \widehat{C}^\dagger_{\bf M} \, \widehat{x} _l \; \left( \widehat{C}_{\bf M} \; \widehat{x}_k \widehat{C}^{-1}_{\bf M}\right)  \widehat{C}_{\bf M} | 0  \rangle = \langle 0 | \widehat{C}^\dagger_{\bf M} \, \widehat{x} _l \; \sum_j \left( {\bf D}_{j k} \, \widehat{x}_j - {\bf B}_{j k} \,\widehat{p}_j \right)  \widehat{C}_{\bf M} | 0  \rangle ,
\end{equation}
\noindent and again, we inserted the expressions for the resulting amplitudes using the covariance matrix components. 

We now plug in (\ref{Amp1}), (\ref{Amp2}) and (\ref{Amp3}) in the relation for the geometric connection (\ref{BerryDef}), and after some simplifications using the block form of matrix ${\bf M}$ and the symplectic group conditions (\ref{Sp1Coord}), we obtain the following form for the geometric phase
\begin{equation}
\gamma_{\bf M} = - \frac{1}{4 \hbar} \int_{\cal C} \mbox{Tr} \left[ \left( \begin{array}{cc} {\bf L}^2 & {\bf 0} \\ {\bf 0} & \hbar^2 {\bf L}^{-2} \end{array}\right) {\bf M}^T \, {\bf \Omega} \, d {\bf M} \right] ,\label{BerryPhaseFormula}
\end{equation}
\noindent where the symbol `Tr'  stands for the trace of the matrix. This expression exhibits an invariance under canonical transformation. To show this, consider a unitary operator $\widehat{C}_{\bf M'}$ related with a new symplectic matrix ${\bf M'}$. The Hamiltonian is given by
\begin{equation}
\widehat{H}' = \widehat{H}_{{ \bf M'} \cdot {\bf M}} = \widehat{C}_{\bf M'} \, \widehat{H}_{\bf M} \widehat{C}^{-1}_{\bf M'} = \widehat{C}_{{\bf M'} \cdot {\bf M}} \, \widehat{H} \widehat{C}^{-1}_{ {\bf M'} \cdot {\bf M}} ,
\end{equation}
\noindent where the group multiplication was used in the second line. The new symplectic matrix ${\bf M'} \cdot {\bf M}$, with ${\bf M'}$ fixed, can be inserted in the geometric phase (\ref{BerryPhaseFormula}) and this yields
\begin{equation}
\gamma_{\bf M} = - \frac{1}{4 \hbar} \int_{\cal C} \mbox{Tr} \left[ \left( \begin{array}{cc} {\bf L}^2 & {\bf 0} \\ {\bf 0} & \hbar^2 {\bf L}^{-2} \end{array}\right) ({\bf M'} \cdot {\bf M})^T \, {\bf \Omega} \, d ({\bf M'} \cdot {\bf M})  \right] = - \frac{1}{4 \hbar} \int_{\cal C} \mbox{Tr} \left[ \left( \begin{array}{cc} {\bf L}^2 & {\bf 0} \\ {\bf 0} & \hbar^2 {\bf L}^{-2} \end{array}\right) {\bf M}^T \cdot {\bf M'}^T \, {\bf \Omega} \, {\bf M'} \cdot d{\bf M}  \right] =\gamma_{\bf M},
\end{equation}
\noindent which shows the invariance of the geometric phase under canonical transformations.

On the other hand, it can be notice that after an integration by parts in (\ref{BerryPhaseFormula}), the boundary term is given by the covariance matrix ${\bf V}^{(2)}$ 
\begin{equation}
\gamma_{\bf M} = \frac{1}{4 \hbar} \int_{\cal C} \mbox{Tr} \left[ {\bf \Omega} \,  {\bf M} \, \left( \begin{array}{cc} {\bf L}^2 & {\bf 0} \\ {\bf 0} & \hbar^2 {\bf L}^{-2} \end{array}\right) \, d {\bf M}^T  \right] - \frac{1}{2\hbar} \int_{\cal C} d \mbox{Tr}\left[ {\Omega} \, {\bf V}^{(2)} \right] ,\label{BerryPhaseFormula2}
\end{equation}
\noindent which is null, when a closed path in the parameters space is considered.

Summarizing, the expression (\ref{BerryPhaseFormula}) (or (\ref{BerryPhaseFormula2})) yields the geometric phase for a variation on the parameters of the group element ${\bf M} \in Sp(2n, \mathbb{R})$. 
This expression is valid for an arbitrary group element ${\bf M}$ and not just for those close to the group identity. Also, from this expression we can read the Berry's connection ${\cal A}_{\bf M}$ for the symplectic group acting on the Gaussian state 
\begin{equation}
\gamma_{\bf M} = \int_{\cal C} {\cal A}_{\bf M}.
\end{equation}
\noindent Due to the group $Sp(2n,\mathbb{R})$ has dimension $n(2n+1)$, there will be $n (2n-1)$ spurious components which can be fixed with a gauge transformation of ${\cal A}_{\bf M}$. However, although a detailed analysis of this connection ${\cal A}_{\bf M}$ is required, it is beyond the scope of this paper and will be studied elsewhere. Instead, let us now show in the following subsections two examples of the geometric phase associated to the squeeze operator, for $n=1$ and $n=2$. Recall that the singular case when ${\bf B} = {\bf 0}$ is treated in the appendix \ref{SingularCase}.

\subsection{Geometric phase for the $n=1$ squeeze operator.}
The squeeze operator for the $n=1$ case is given by
\begin{equation}
\widehat{S}^{(1)}(\zeta) = e^{\frac{1}{2} ( \zeta^* \widehat{a}^2 - \zeta (\widehat{a}^\dagger)^2 )},
\end{equation}
\noindent where $\widehat{a}$ and $\widehat{a}^\dagger $ are the ladder operators satisfying the canonical commutation relation $[\widehat{a}, \widehat{a}^\dagger] = 1$. The real parameter $r$, which is the absolute value of the parameter $\zeta = r e^{i \theta} \in \mathbb{C}$, labels the squeezing degree of the system. We now use the transformation $\widehat{a} = \frac{1}{\sqrt{2}} \frac{\widehat{x}}{l} + \frac{i}{\sqrt{2}} \frac{l \, \widehat{p}}{\hbar}$, to obtain the squeeze operator in the Schr\"odinger representation
\begin{equation}
\widehat{S}^{(1)}(\zeta) = e^{- \frac{i}{2 \hbar} \left( \begin{array}{cc} \widehat{x} & \widehat{p} \end{array} \right) \left( \begin{array}{cc} \frac{\hbar r}{l^2} \sin \theta & - r \cos \theta \\ - r \cos \theta & - \frac{l^2 \, r}{\hbar} \sin \theta \end{array} \right) \left( \begin{array}{c} \widehat{x} \\ \widehat{p} \end{array}\right) }.\label{SqueezeOp}
\end{equation}

Here, the Lie algebra isomorphism between $sp(2,\mathbb{R})$ and the Lie algebra of the second order polynomials \cite{de2011symplectic, hall2018theory} allows us to take the matrix
\begin{equation}
{\bf L}^{(1)}_s = \left( \begin{array}{cc} \frac{\hbar}{l^2} r \sin \theta & - r \cos \theta \\ - r \cos \theta & - \frac{l^2}{\hbar} r \sin \theta \end{array} \right),
\end{equation}
\noindent as the Lie algebra element of $sp(2,\mathbb{R})$. To make a difference between this description and the $n=2$ case further below, we introduce the index ${}^{(1)}$, which indicates we are working with $n=1$. In both cases, ${\bf L}^{(j)}_s$ with $j=1$ or $j=2$, corresponds to the Lie algebra element in $sp(2j,\mathbb{R})$ arising from the corresponding squeeze operator $\widehat{S}^{(j)}(\zeta)$.

The symplectic matrix ${\bf M}_s({\bf L}^{(1)}_s)$ associated to this operator is given by the exponential map of the matrix ${\bf L}^{(1)}_s$ which yields
\begin{equation}
{\bf M}_s({\bf L}^{(1)}_s) = \left( \begin{array}{cc} \cosh r - \cos \theta \sinh r & - \frac{l^2}{\hbar}\sin \theta \sinh r \\ - \frac{\hbar}{l^2} \sin \theta \sinh r & \cosh r + \cos \theta \sinh r \end{array}  \right).
\end{equation}

We now insert this matrix in (\ref{BerryPhaseFormula}) and consider the trajectory ${\cal C}$ given by a constant $r=R$ and $\theta = [0, 2 \pi)$. A quick calculation gives the following expression for the corresponding geometric phase
\begin{equation}
\gamma^{(1)}_{\bf s} = - \pi \sinh^2 (R), \label{Berryn1}
\end{equation}
\noindent which coincides with the same calculation reported in \cite{chaturvedi1987berry}. Remarkably, despite the factor ${\frac{1}{\hbar}}$ in (\ref{BerryPhaseFormula}) and the factors $\hbar/l^2$ and $l^2/\hbar$ in the squeeze operator, the geometric phase is $\hbar$ and $l^2$ independent. It only depends on the squee parameter $r$. We might expect that this feature is intrinsic of the system with $n=1$, but, as we will soon notice, is also present in the $n=2$ case. 

\subsection{Geometric phase for the $n=2$ squeeze operator.}

The squeeze operator $\widehat{S}^{(2)}(\zeta)$ for a bi-partite system is of the form
\begin{equation}
\widehat{S}^{(2)}(\zeta) = e^{ \left( \zeta^* \widehat{a}_1 \widehat{a}_2 - \zeta \widehat{a}^\dagger_1 \widehat{a}^\dagger_2 \right) }, \label{SqOperator}
\end{equation}
\noindent where as before, $\widehat{a}_1$ and $\widehat{a}_2$ are the annihilation operators for the sub-systems, say, 1 and 2, of the bi-partite system. $\widehat{a}^\dagger_1$ and $\widehat{a}^\dagger_2$ are their adjoint operators respectively and $\zeta = r e^{i \phi}$ is a complex number labelling the amount of squeezing. 

Again, the operator in  (\ref{SqOperator}) is given in the Fock representation and we write it in the Schr\"odinger representation using the transformation
\begin{eqnarray}
\widehat{a}_j =  \frac{1}{\sqrt{2}} \frac{\widehat{x}_j}{l_j} + \frac{i  }{\sqrt{2}}  \frac{ l_j \widehat{p}_j}{\hbar} , \qquad \widehat{a}^\dagger_j =  \frac{1}{\sqrt{2}} \frac{\widehat{x}_j}{l_j} - \frac{i  }{\sqrt{2}}  \frac{ l_j \widehat{p}_j}{\hbar} , \qquad j=1,2.
\end{eqnarray}
\noindent  Once we insert these expressions for $\widehat{a}_j$ and $\widehat{a}^\dagger_j$ in the expression for $\widehat{S}^{(2)}(\zeta)$, it takes the following form
\begin{eqnarray}
\widehat{S}^{(2)}(\zeta) &=& \exp \left\{ \frac{-i}{\hbar} \left[  \frac{\hbar \, \zeta_y}{l_1 l_2} \; \widehat{x}_1 \widehat{x}_2 -  \frac{l_2 \, \zeta_x}{l_1} \; \widehat{x}_1 \widehat{p}_2 -  \frac{l_1\, \zeta_x}{l_2} \; \widehat{p}_1 \widehat{x}_2  -  \frac{l_1 l_2 \, \zeta_y}{\hbar} \widehat{p}_1 \widehat{p}_2  \right] \right\}, \label{SqGen}
\end{eqnarray}
\noindent where $\zeta_x$ and $\zeta_y$ are the real and imaginary parts of $\zeta$. The operators in the argument of (\ref{SqGen}) can be accommodated in a matrix form as follows
\begin{equation}
\widehat{S}^{(2)}(\zeta) = \exp\left\{ - \frac{i}{2\hbar} (\vec{\widehat{R}}^T_1, \vec{\widehat{R}}^T_2) \left( \begin{array}{cc} {\bf 0} & {\bf b} \\ {\bf b}^T & {\bf 0} \end{array}\right) \left( \begin{array}{c} \vec{\widehat{R}}_1 \\ \vec{\widehat{R}}_2 \end{array}\right) \right\}. \label{SqueezedGen}
\end{equation}
\noindent where ${\bf b}$ is given by
\begin{equation}
{\bf b} = \left( \begin{array}{cc} \frac{\hbar}{l_1 l_2} \zeta_y & -\frac{l_2}{l_1} \zeta_x \\ -\frac{l_1}{l_2}  \zeta_x & -\frac{l_1 l_2}{\hbar} \zeta_y  \end{array}\right). \label{DimensionFullb}
\end{equation}

The Lie algebra isomorphism between $sp(4,\mathbb{R})$ and the Lie algebra of the second order polynomials \cite{de2011symplectic, hall2018theory}, allows us to take the $4 \times 4$ square matrix in (\ref{SqueezedGen}) to be the Lie algebra element ${\bf L}^{(2)}_s \in sp(4,\mathbb{R})$. This matrix ${\bf L}^{(2)}_s$ has the block matrices ${\bf a} = {\bf c} = {\bf 0}$ (see the appendix \ref{LAAnalysis} for more details) and ${\bf b}$ is given in (\ref{DimensionFullb}).

 The matrix ${\bf L}^{(2)}_s$ is now inserted in the formula (\ref{FormulaForSQ}) of Appendix \ref{LAAnalysis} yielding the expression for $\widetilde{\bf M}$ which, after using the transformation $\Gamma$ in (\ref{RelationbetweenMs}) gives the following form for the matrix ${\bf M}({\bf L}^{(2)}_s)$ 
\begin{equation}
{\bf M}({\bf L}^{(2)}_s) = \left( \begin{array}{cccc} \cosh(r) & - \frac{l_1}{l_2} \sinh(r) \cos\phi & 0 & - \frac{l_1 l_2}{\hbar} \sinh(r) \sin\phi \\  - \frac{l_2}{l_1} \sinh(r) \cos\phi & \cosh(r) & - \frac{l_1 l_2}{\hbar} \sinh(r) \sin\phi & 0 \\  0 & - \frac{\hbar}{l_1 l_2} \sinh(r) \sin\phi & \cosh(r) &  \frac{l_2}{l_1} \sinh(r) \cos\phi \\ - \frac{\hbar}{l_1 l_2} \sinh(r) \sin\phi & 0 &  \frac{l_1}{l_2} \sinh(r) \sin\phi & \cosh(r)\end{array}\right). \label{MforN2}
\end{equation}

Let us now consider the same trajectory used for the $n=1$ case, that is to say, $r = R$ is constant and $\phi \in [0, 2 \pi)$ and insert this matrix in (\ref{BerryPhaseFormula}). After some matrix calculations and the corresponding integration in $\phi$ it yields the following result
\begin{equation}
\gamma^{(2)}_{\bf s} = - 2\pi \sinh^2 (R). \label{Berryn2}
\end{equation}
\noindent As can be notice, this value for the geometric phase doubles the geometric phase for the $n=1$ case using the same trajectory in the parameters space. This implies that for the same amount of squeezing $r$ the observed geometric phase for $n=2$ must double that of the $n=1$ case. Remarkably, both geometric phases (\ref{Berryn1}) and (\ref{Berryn2}) are $\hbar$ and $l_j$ independent and depend only on the squeezing parameter $r$.

\section{Conclusions} \label{Discussion}
 
The main results of this work are the expressions (\ref{BerryPhaseFormula}) and (\ref{BerryPhaseFormula2}) for the geometric phases of a general Gaussian state of the form $\Psi_{\bf M}$ and an arbitrary path ${\cal C}$. This Gaussian state corresponds to a state generated by the action of the operator $\widehat{C}_{\bf M}$ on the vacuum state $\Psi_0(\vec{x})$ for a $n$-partite system. The operator $\widehat{C}_{\bf M}$ is the unitary representation of the symplectic group in the Hilbert space ${\cal H}$ given in section \ref{CovarianceMatrix}.

Our first observation is that, despite the ${\bf B}^{-1}$ term in the amplitudes  (\ref{Amp1}), (\ref{Amp2}) and (\ref{Amp3}), we managed to re-write the geometric phase in a presentable way and also, we showed how the covariance matrix ${\bf V}^{(2)}$ arises as a boundary term in (\ref{BerryPhaseFormula2}). The special case for ${\bf B}= {\bf 0}$ was considered in the appendix (\ref{SingularCase}). We also showed the invariance of the geometric phase under canonical transformations.

Surprisingly, modulo other constant matrices, the Berry connection ${\cal A}_{\bf M}$ results to be a `bilinear' term in the symplectic group matrices, that is to say, ${\cal A}_{\bf M} \sim {\bf M}^T \, {\Omega} \, d{\bf M}$ . Therefore, this result paves the way to explore the Berry curvature and its singular points.
 
Additionally, we calculated the geometric phase for the squeeze operators $\widehat{S}^{(j)}(\zeta)$ with $j=1,2$ and showed, in the case of $n=2$, that the geometric phase is twice the geometric phase for $n=1$. Moreover, both geometric phases are $\hbar$ and $l_j$ independent. This feature might induce some sort of `classical nature' to these phases. However, due to the squeezed states are quantum states with no classical analog, the quantum nature of the their corresponding geometric phase is granted. The absence of dimension-full parameters like $\hbar$ or $l_j$ in the squeeze phase suggest that the dynamical of the system does not play a relevant role in these phases, i.e., their nature is more like a kinematical feature of the symplectic group topology \cite{mukunda2014classical}. 
  
Finally, it is worth to mention that in order to derive the expression for the geometric phase, we first derived the covariance matrix for Gaussian states in (\ref{CVMatrix}), and we also provided the full relation between $sp(4,\mathbb{R})$ and $Sp(4,\mathbb{R})$ in the appendix \ref{LAAnalysis}. This result showed in the appendix \ref{LAAnalysis}, to the best of the author's knowledge, have not been reported elsewhere. The full relation between the symplectic group $Sp(4,\mathbb{R})$ and its Lie algebra can be used to explore different geometric phases and not just those related with the squeeze operator.

\section{Acknowledgments}
I thank Andrea Mari, Alessandro Bravetti and M. Berm\'udez-Monta\~na for their comments and helpful discussions.

\section{Appendix: Relation between $Sp(4,\mathbb{R})$ and $sp(4,\mathbb{R})$} \label{LAAnalysis}

In this appendix we derive the direct relation between the Lie algebra elements $sp(4,\mathbb{R})$ and the symplectic group matrices $Sp(4,\mathbb{R})$.

\subsection{Lie algebra and group analysis} 



Consider an arbitrary element ${\bf m} \in sp(4,\mathbb{R})$ of the form
\begin{equation}
{\bf m} = \left( \begin{array}{cc} {\bf J} & 0 \\ 0 & {\bf J} \end{array} \right) {\bf L}, \label{LieAlegebraElem}
\end{equation}
\noindent where the matrix ${\bf L}$ is a real symmetric matrix written as
\begin{equation}
{\bf L}= \left( \begin{array}{cc} {\bf a} & {\bf b} \\ {\bf b}^T & {\bf c} \end{array}\right), \label{LMatrix}
\end{equation}
\noindent and ${\bf a}$ and ${\bf c}$ are $2\times2$ symmetric matrices and ${\bf b}$ is a $2\times2$ matrix. 

The elements of the symplectic group $Sp(4, \mathbb{R})$ close to the identity can be obtained via the exponential map \cite{hall2018theory} of the Lie algebra element ${\bf L}$ as
\begin{equation}
\widetilde{\bf M} = \exp{ \left[ \left( \begin{array}{cc} {\bf J} & 0 \\ 0 & {\bf J} \end{array} \right) {\bf L} \right]} = \left( \begin{array}{cc} \widetilde{\bf A} & \widetilde{\bf B} \\ \widetilde{\bf C} & \widetilde{\bf D} \end{array}\right).  \label{MandS}
\end{equation}
\noindent The aim of this section is to obtain the relation between the block matrices $\widetilde{\bf A}$, $\widetilde{\bf B}$, $\widetilde{\bf C}$ and $\widetilde{\bf D}$ and the Lie algebra element ${\bf L}$. What we will obtain is a relation between the block matrices ${\bf a}$, ${\bf b}$ and ${\bf c}$ of the Lie algebra element ${\bf L}$ and the matrices  $\widetilde{\bf A}$, $\widetilde{\bf B}$, $\widetilde{\bf C}$ and $\widetilde{\bf D}$ of the group element $\widetilde{\bf M}$. Recall that we are using a group action which is different, although equivalent, to the one used for the calculation of the covariance matrix and consequently, for the geometric phase, that is to say, in this section we use $\widetilde{\bf M}$ instead of ${\bf M}$.

To proceed, let us expand the exponential in (\ref{MandS}) and collect together the even and odd terms of the expansion as follows
\begin{widetext}
\begin{eqnarray}
\widetilde{\bf M} &=& \left[  {\bf 1} + \frac{1}{2!} {\bf S} + \dots + \frac{1}{(2n)!} {\bf S}^n + \dots 	\right] +  \sqrt{{\bf S}} \left[ {\bf 1} + \frac{1}{3!}  {\bf S} + \dots + \frac{1}{(2n+1)!} {\bf S}^n + \dots   \right], \label{Expansion} 
\end{eqnarray}
\end{widetext}
\noindent where the matrix ${\bf S}$ is defined as
\begin{widetext}
\begin{equation}
{\bf S} = \left[ \left( \begin{array}{cc} {\bf J} & 0 \\ 0 & {\bf J} \end{array} \right) \left( \begin{array}{cc} {\bf a} & {\bf b} \\ {\bf b}^T & {\bf c} \end{array} \right) \right]^{2} = \left( \begin{array}{cc} - (\det {\bf a} + \det {\bf b}) {\bf 1}_{2\times2} & {\bf J} {\bf d} \\ - {\bf J} {\bf d}^T & - (\det {\bf b} + \det {\bf c}) {\bf 1}_{2\times2} \end{array}\right), \label{SDefinition}
\end{equation}
\end{widetext}
\noindent and the matrix ${\bf d}$ is given by ${\bf d} = {\bf a} {\bf J} {\bf b} + {\bf b} {\bf J} {\bf c}$. The notation used in (\ref{Expansion}) for $\sqrt{\bf S}$ refers to the matrix
\begin{equation}
\sqrt{\bf S} := \left( \begin{array}{cc} {\bf J} & 0 \\ 0 & {\bf J} \end{array} \right) \left( \begin{array}{cc} {\bf a} & {\bf b} \\ {\bf b}^T & {\bf c} \end{array} \right).
\end{equation}

As can be seen from the expansion (\ref{Expansion}), in order to obtain the expression for $\widetilde{\bf M}$ we need first to determine ${\bf S}^n$ and then, we have to insert the expression for ${\bf S}^n$ in (\ref{Expansion}) and calculate both sums therein. Let us  proceed in the next subsection with the first step: the calculation of ${\bf S}^{n}$. 

\subsection{Calculation of ${\bf S}^{n}$}

The matrix ${\bf S}$ is formed by four $2 \times 2$ block matrices where the upper left and the lower right are multiples of the identity matrix ${\bf 1}_{2\times2}$. The upper right block is the matrix ${\bf J} {\bf d}$ whereas the lower left is $- {\bf J} {\bf d}^T$. Notably, we found that this block structure is preserved after exponentiating the matrix ${\bf S}$ an integer number of times. That is to say, the n-power of matrix ${\bf S}$ yields a new matrix ${\bf S}^n$ given as
\begin{equation}
{\bf S}^n = \left( \begin{array}{cc} \alpha_n{\bf 1}_{2\times2} & \beta_n {\bf J} {\bf d} \\ - \beta_n {\bf J} {\bf d}^T & \gamma_n {\bf 1}_{2\times2} \end{array}\right). \label{MatrixSDef}
\end{equation} 
\noindent The coefficients $\alpha_n$, $\beta_n$ and $\gamma_n$, to be determined, depend on the values of the matrices ${\bf a}$, ${\bf b}$, ${\bf c}$ and ${\bf d}$. For $n=1$, these coefficients are given by the factors in the block matrices of ${\bf S} $ in  (\ref{SDefinition}) and can be directly defined as
\begin{equation}
\alpha_1 := - (\det {\bf a} + \det {\bf b}), \; \beta_1 := +1, \; \gamma_1 := - (\det {\bf c} + \det {\bf b}). \label{initialvalues}
\end{equation}

To calculate these coefficients $\alpha_n$, $\beta_n$ and $\gamma_n$ for arbitrary $n$, first note that they can be generated with a linear operator ${\bf T}$ as
\begin{eqnarray}
\left( \begin{array}{c} \alpha_n \\ \beta_n \\ \gamma_n \end{array}\right) = {\bf T}^{n-1} \left( \begin{array}{c} \alpha_1 \\ \beta_1 \\ \gamma_1 \end{array}\right), \label{EqForCoeff}
\end{eqnarray}
\noindent where the matrix ${\bf T}$ is given by
\begin{equation}
{\bf T} = \left( \begin{array}{ccc} \alpha_1 & \beta_1 \det {\bf d} & 0 \\ \beta_1 & \gamma_1 & 0 \\ 0 & \beta_1 \det {\bf d} & \gamma_1 \end{array}\right).
\end{equation}

The calculation shows that the $n-1$ power of ${\bf T}$ is a matrix of the form
\begin{equation}
{\bf T}^{n-1} = \left( \begin{array}{cc} {\bf U}^{n-1} & \vec{0}^T \\ \vec{u}^T \gamma_1^{n-2} \sum^{n-2}_{j=0} \gamma^{-j}_1 {\bf U}^j & \gamma^{n-1}_1 \end{array}\right),
\end{equation}
\noindent where $\vec{0} = (0,0)$ and $\vec{u} = (0,  \beta_1 \det d)$ and matrix ${\bf U}$ is given by
\begin{eqnarray}
{\bf U} = \left( \begin{array}{cc} \alpha_1 & \beta_1 \det {\bf d} \\ \beta_1 & \gamma_1\end{array}\right).
\end{eqnarray}

Then, using (\ref{EqForCoeff}) we have the following relation for the coefficients
\begin{eqnarray}
\left( \begin{array}{c} \alpha_n \\ \beta_n \end{array}\right) &=& {\bf U}^{n-1} \left( \begin{array}{c} \alpha_1 \\ \beta_1 \end{array}\right), \\
\gamma_n &=& \gamma^n_1 + \vec{u}^T \gamma^{n-2}_1 \sum^{n-2}_{j=0} \gamma^{-j}_1 {\bf U}^j \left( \begin{array}{c} \alpha_1 \\ \beta_1 \end{array}\right). \label{FExpressionCoeff}
\end{eqnarray}

In order to calculate ${\bf U}^{n-1}$ we need to diagonalize matrix ${\bf U}$ hence, let ${\bf P}$ be the matrix diagonalizing ${\bf U}$, then 
\begin{equation}
{\bf U} = {\bf P} \, {\bf U}_d \, {\bf P}^{-1}, \label{EqForU}
\end{equation}
\noindent where ${\bf U}_d$ is the diagonal matrix and the matrix ${\bf P}$ is
\begin{equation}
{\bf P} = \left( \begin{array}{cc} \frac{(\lambda_+ - \gamma_1)}{\beta_1 } k_1 & \frac{(\lambda_{-} - \gamma_1)}{\beta_1} k_2 \\ k_1 & k_2 \end{array}\right).
\end{equation}

The real arbitrary parameters $k_1$ and $k_2$ result from the diagonalization procedure. Its values will be automatically cancelled as part of the calculation of ${\bf U}^{n-1}$ further below. The eigenvalues of ${\bf U}$, denoted by $\lambda_{\pm}$, have the following expressions 
\begin{widetext}
\begin{eqnarray}
\lambda_{\pm} &=& - \frac{\det {\bf a}  + \det {\bf c} + 2 \det {\bf b} }{2} \pm \frac{1}{2} \sqrt{ (\det {\bf a} - \det {\bf c})^2 + 4  \det {\bf d} }, \label{eigenvalues}
\end{eqnarray}
\end{widetext}
\noindent and the diagonal matrix ${\bf U}_d$ is
\begin{equation}
{\bf U}_d = \left( \begin{array}{cc} \lambda_+ & 0 \\ 0 & \lambda_{-} \end{array}\right).
\end{equation}

We now take the $n-1$ power of ${\bf U}$ given in (\ref{EqForU}) to obtain the following result
\begin{widetext}
\begin{eqnarray}
{\bf U}^{n-1} = \left( \begin{array}{cc} \frac{(\lambda_+ - \gamma_1)}{\beta_1 } k_1 & \frac{(\lambda_{-} - \gamma_1)}{\beta_1} k_2 \\ k_1 & k_2 \end{array}\right) \left( \begin{array}{cc} \lambda^{n-1}_+ & 0 \\ 0 & \lambda^{n-1}_{-} \end{array}\right) \left( \begin{array}{cc} \frac{(\lambda_+ - \gamma_1)}{\beta_1 } k_1 & \frac{(\lambda_{-} - \gamma_1)}{\beta_1} k_2 \\ k_1 & k_2 \end{array}\right)^{-1},
\end{eqnarray}
\end{widetext}
\noindent which, when combined with the result in (\ref{FExpressionCoeff}) together with the expression for $\vec{u}$, gives
\begin{eqnarray}
\alpha_n &=& \frac{\left[ ( \lambda_+ - \gamma_1 ) \lambda^n_+ - ( \lambda_- - \gamma_1 ) \lambda^n_-   \right]}{\sqrt{ (\alpha_1 - \gamma_1)^2 + 4 \beta^2_1 \det {\bf d} } }  ,\label{alpha} \\
\beta_n &=& \frac{\left[ \lambda^n_+ -  \lambda^n_-   \right]}{\sqrt{ (\alpha_1 - \gamma_1)^2 + 4 \beta^2_1 \det {\bf d} } }  ,\label{beta} \\
\gamma_n &=& \frac{\left[ ( \lambda_+ - \gamma_1 ) \lambda^n_- - ( \lambda_- - \gamma_1 ) \lambda^n_+   \right]}{\sqrt{ (\alpha_1 - \gamma_1)^2 + 4 \beta^2_1 \det {\bf d} } }. \label{gamma}
\end{eqnarray}
\noindent These are the final expressions for the coefficients in ${\bf S}^n$. We are now ready to move to the second step: the analysis of the infinite series in (\ref{Expansion}).

\subsection{Series analysis}

Using the expression for the $n$ power of matrix ${\bf S}$, defined in (\ref{MatrixSDef}), the expression (\ref{Expansion}) can be written as
\begin{widetext}
\begin{eqnarray}
\widetilde{\bf M} &=& {\bf 1} + \sum^{+\infty}_{n=1} \frac{1}{(2n)!} \left( \begin{array}{cc} \alpha_n & \beta_n {\bf J} {\bf d} \\ - \beta_n {\bf J} {\bf d}^T & \gamma_n \end{array}\right) +  \sqrt{{\bf S}}\left[ {\bf 1} + \sum^{+\infty}_{n=1} \frac{1}{(2n+1)!}  \left( \begin{array}{cc} \alpha_n & \beta_n {\bf J} {\bf d} \\ - \beta_n {\bf J} {\bf d}^T & \gamma_n \end{array}\right) \right]. \label{GroupMatrix}
\end{eqnarray}
\end{widetext}
\noindent After collecting the components of each block matrix we obtain the following coefficients
\begin{widetext}
\begin{eqnarray}
\alpha^{(e)} := 1 + \sum^{+\infty}_{n=1} \frac{1}{(2n)!} \alpha_n, \qquad \beta^{(e)} := \sum^{+\infty}_{n=1} \frac{1}{(2n)!} \beta_n, \qquad \gamma^{(e)} := 1 + \sum^{+\infty}_{n=1} \frac{1}{(2n)!} \gamma_n, \label{EvenCoeff}\\
\alpha^{(o)} := 1 + \sum^{+\infty}_{n=1} \frac{1}{(2n+1)!} \alpha_n, \qquad \beta^{(o)} := \sum^{+\infty}_{n=1} \frac{1}{(2n+1)!} \beta_n, \qquad \gamma^{(o)} := 1 + \sum^{+\infty}_{n=1} \frac{1}{(2n+1)!}\gamma_n. \label{OddCoeff}
\end{eqnarray}
\end{widetext}
We now insert (\ref{alpha}), (\ref{beta}) and (\ref{gamma}) in the relations (\ref{EvenCoeff}) - (\ref{OddCoeff}) to obtain the final form of the coefficients
\begin{widetext}
\begin{eqnarray}
 \alpha^{(e)} &=&  \frac{ (\lambda_+ + \det{\bf b} + \det{\bf c}) \cosh\sqrt{\lambda_+} - (\lambda_- + \det{\bf b} + \det{\bf c}) \cosh\sqrt{\lambda_- } }{\sqrt{ (\det{\bf a} - \det{\bf c})^2 + 4 \det {\bf d} } } , \label{alphapar} \\
 \alpha^{(o)} &=& \frac{ (\lambda_+ + \det{\bf b} + \det{\bf c}) \frac{ \sinh\sqrt{\lambda_+}}{\sqrt{\lambda_+}}  - (\lambda_- + \det{\bf b} + \det{\bf c}) \frac{ \sinh\sqrt{\lambda_- }}{\sqrt{\lambda_-}}  }{\sqrt{ (\det{\bf a} - \det{\bf c})^2 + 4 \det {\bf d}  } } , \label{alphaimpar} \\
\beta^{(e)} &=& \frac{  \cosh\sqrt{\lambda_+} -  \cosh\sqrt{\lambda_- }}{\sqrt{ (\det{\bf a} - \det{\bf c})^2 + 4 \det {\bf d}  } } , \label{betapar} \\
\beta^{(o)} &=& \frac{\frac{ \sinh\sqrt{\lambda_+} }{\sqrt{\lambda_+}} - \frac{ \sinh\sqrt{\lambda_- }}{\sqrt{\lambda_-}} }{\sqrt{ (\det{\bf a} - \det{\bf c})^2 + 4 \det {\bf d}  } } , \label{betaimpar} \\
 \gamma^{(e)} &=& \frac{ (\lambda_+ + \det{\bf b} + \det{\bf c}) \cosh\sqrt{\lambda_-} - (\lambda_- + \det{\bf b} + \det{\bf c}) \cosh\sqrt{\lambda_+ } }{\sqrt{ (\det{\bf a} - \det{\bf c})^2 + 4 \det {\bf d}  } } , \label{gammapar} \\
 \gamma^{(o)} &=& \frac{ (\lambda_+ + \det{\bf b} + \det{\bf c}) \frac{ \sinh\sqrt{\lambda_-}}{\sqrt{\lambda_-}} - (\lambda_- + \det{\bf b} + \det{\bf c}) \frac{\sinh\sqrt{\lambda_+ }}{\sqrt{\lambda_+}} }{\sqrt{(\det{\bf a} - \det{\bf c})^2 + 4 \det {\bf d}  } } , \label{gammaimpar}
\end{eqnarray}
\end{widetext}
\noindent where we have to recall the expressions for the eigenvalues $\lambda_\pm$ in (\ref{eigenvalues}). We now use these results to provide the final expression for the matrix components in (\ref{GroupMatrix}) 
\begin{widetext}
\begin{eqnarray}
\widetilde{\bf A}({\bf a}, {\bf b}, {\bf c}) &=& \alpha^{(e)} + (\alpha^{(o)} - \beta^{(o)} \det{\bf b}) \, {\bf J} \, {\bf a} + \beta^{(o)} {\bf J} \,{\bf b} \, {\bf J} \, {\bf c}\, {\bf J} \, {\bf b}^T, \label{TerminoA}\\
\widetilde{\bf B}({\bf a}, {\bf b}, {\bf c}) &=& (\gamma^{(o)} - \beta^{(o)} \det{\bf a}) \, {\bf J} \, {\bf b} + \beta^{(e)} ( {\bf J} \,{\bf a} \, {\bf J} \, {\bf b} +  {\bf J} \,{\bf b} \, {\bf J} \, {\bf c} )+  \beta^{(o)}  {\bf J} \,{\bf a} \, {\bf J} \, {\bf b} \, {\bf J} \, {\bf c}, \label{TerminoB} \\
\widetilde{\bf C}({\bf a}, {\bf b}, {\bf c}) &=& (\alpha^{(o)} - \beta^{(o)} \det{\bf c}) \, {\bf J} \, {\bf b}^T + \beta^{(e)} ( {\bf J} \,{\bf b}^T \, {\bf J} \, {\bf a} +  {\bf J} \,{\bf c} \, {\bf J} \, {\bf b}^T )+  \beta^{(o)}  {\bf J} \,{\bf c} \, {\bf J} \, {\bf b}^T \, {\bf J} \, {\bf a}, \label{TerminoC} \\
\widetilde{\bf D}({\bf a}, {\bf b}, {\bf c}) &=& \gamma^{(e)} + (\gamma^{(o)} - \beta^{(o)} \det{\bf b}) \, {\bf J} \, {\bf c} + \beta^{(o)} {\bf J} \,{\bf b}^T \, {\bf J} \, {\bf a}\, {\bf J} \, {\bf b}. \label{TerminoD}
\end{eqnarray}
\end{widetext}

These expressions provide the relation between the components ${\bf a}$, ${\bf b}$ and ${\bf c}$ of the Lie algebra element ${\bf m}$ in (\ref{LieAlegebraElem}) with the corresponding symplectic matrix $\widetilde{\bf M}$. To the best of the author's knowledge, this result is new and have not been reported elsewhere.

A particular case is when ${\bf a} = {\bf c} = 0$ which is connected with the form of the squeeze operators in section \ref{BerryPhase}. Inserting these values on the previous expressions yield 

\begin{equation}
\widetilde{\bf M}({\bf 0}, {\bf b}, {\bf 0}) =  \left( \begin{array}{cc} \cosh(\sqrt{- \det{\bf b}})  & \frac{\sinh(\sqrt{- \det{\bf b}})}{\sqrt{- \det{\bf b}}} {\bf J} {\bf b} \\ \frac{\sinh( \sqrt{- \det{\bf b}})}{\sqrt{- \det{\bf b}}} {\bf J} {\bf b}^T  & \cosh^2(\sqrt{- \det{\bf b}})   \end{array}\right) . \label{FormulaForSQ}
\end{equation}



\section{Appendix: geometric phase for ${\bf B} = {\bf 0}$.} \label{SingularCase}

In this appendix we will briefly consider the special case when the symplectic matrix ${\bf M}$ has the block matrix ${\bf B} = {\bf 0}$. In this case, the block form of ${\bf M}$ is given as
\begin{equation}
{\bf M}({\bf B} = {\bf 0}) = \left( \begin{array}{cc} {\bf A} & {\bf 0} \\ {\bf C} & {\bf A}^{- T} \end{array} \right), \label{ScaseB0}
\end{equation}
 the representation of the operator $\widehat{C}_{\bf M}$ is now given as
\begin{equation}
\widehat{C}_{\bf M} \Psi(\vec{x}) = \frac{1}{\det {\bf A}} e^{\frac{i}{2 \hbar} \vec{x}^T {\bf C} {\bf A}^{-1} \vec{x} } \Psi({\bf A}^{-1} \vec{x}), \label{SingularB}
\end{equation}
\noindent hence it only depends on the matrices ${\bf A}$ and ${\bf C}$. The geometric phase in this case can be written as
\begin{equation}
\gamma_{\bf M} = i \int_{\cal C} \left[ \langle 0 | \widehat{C}^{\dagger}_{\bf M} \frac{\partial }{\partial A_{ab}} \widehat{C}_{\bf M} | 0 \rangle d A_{a b} + \langle 0 | \widehat{C}^{\dagger}_{\bf M} \frac{\partial }{\partial C_{ab}} \widehat{C}_{\bf M} | 0 \rangle d C_{a b} \right], \label{BerryB0}
\end{equation}
\noindent where $A_{ab}$ and $C_{ab}$ are the components of the matrices ${\bf A}$ and ${\bf C}$ respectively. We now use the representation in (\ref{SingularB}) and calculate the amplitudes in (\ref{BerryB0})
\begin{eqnarray}
\langle 0 | \widehat{C}^{\dagger}_{\bf M} \frac{\partial }{\partial A_{ab}} \widehat{C}_{\bf M} | 0 \rangle &=& - \frac{i}{4\hbar} l^2_b A_{j b} C_{j k} A^{-1}_{k a} ,   \\
\langle 0 | \widehat{C}^{\dagger}_{\bf M} \frac{\partial }{\partial C_{ab}} \widehat{C}_{\bf M} | 0 \rangle &=& \frac{i}{4\hbar} l^2_b A_{a b},
\end{eqnarray}
\noindent which when inserted in (\ref{BerryB0}) gives
\begin{equation}
\gamma_{\bf M} = \frac{1}{4 \hbar} \int_{\cal C} \mbox{Tr}\left[ {\bf L}^2 {\bf A}^T {\bf C} {\bf A}^{-1} d{\bf A} - {\bf L}^2 {\bf A}^T d{\bf C} \right] = - \frac{1}{4 \hbar} \int_{\cal C} \mbox{Tr}\left[ {\bf L}^2 \, {\bf A}^T \, d{\bf C} - {\bf L}^2 \, {\bf C}^T \, d{\bf A} \right].
\end{equation}
\noindent This result coincides with (\ref{BerryPhaseFormula}) for a symplectic matrix of the form given by (\ref{ScaseB0}). It can be notice that when the block matrix ${\bf C}$ is null, the phase in (\ref{SingularB}) is zero and consequently, the geometric phase is also zero independently of the expression for ${\bf A}$.


\begin{thebibliography}{99}

\bibitem{bennett2000quantum} Bennett, C. H. and DiVincenzo, D. P., ``Quantum information and computation'', \emph{nature}, {\bf 404}, 6775, 2000.

\bibitem{divincenzo1995quantum} DiVincenzo, D. P., ``Quantum computation'', \emph{Science}, {\bf 270}, 5234, 1995.

\bibitem{pan2001entanglement} Pan, J., Simon, C., Brukner, {\v{C}}. and Zeilinger, A., ``Entanglement purification for quantum communication'', \emph{Nature}, {\bf 410}, 6832, 2001.

\bibitem{axline2018demand} Axline, Christopher and others., ``On-demand quantum state transfer and entanglement between remote microwave cavity memories'', \emph{Nature Physics}, {\bf 14}, 7, 2018.

\bibitem{bouwmeester1997experimental} Bouwmeester, D., Pan, J., Mattle, K., Eibl, M., Weinfurter, H. and Zeilinger, A., ``Experimental quantum teleportation'', \emph{Nature}, {\bf 390}, 6660, 1997.

\bibitem{adesso2007entanglement} Adesso, G. and Illuminati, F., ``Entanglement in continuous-variable systems: recent advances and current perspectives'', \emph{Journal of Physics A: Mathematical and Theoretical}, {\bf 40}, 28, 2007.

\bibitem{horodecki2009quantum} Horodecki, R., Horodecki, P., Horodecki, M. and Horodecki, K, ``Quantum entanglement'', \emph{Reviews of modern physics}, {\bf 81}, 2, 2009.

\bibitem{richens2017entanglement} Richens, J. G., Selby, J. H. and Al-Safi, S. W., ``Entanglement is necessary for emergent classicality in all physical theories'', \emph{Physical review letters}, {\bf 119}, 8, 2017.


\bibitem{peres1996separability}   Peres, A., ``Separability criterion for density matrices'', \emph{Physical Review Letters}, {\bf 77}, 8, 1996.

\bibitem{horodecki1997separability}  Horodecki, P., \emph{Separability criterion and inseparable mixed states with positive partial transposition}, Physics Letters A, {\bf 232}, 5, 1997.

\bibitem{simon2000peres} Simon, R., ``Peres-Horodecki separability criterion for continuous variable systems'', \emph{Physical Review Letters}, {\bf 84}, 12, 2000.

\bibitem{duan2000inseparability} Duan, L., Giedke, G., Cirac, J. I. and Zoller, P., \emph{Inseparability criterion for continuous variable systems}, Physical Review Letters, {\bf 84}, 12, 2000.

\bibitem{werner2001bound} Werner, R. F. and Wolf, M. M., \emph{Bound entangled Gaussian states}, Physical review letters, {\bf 86}, 16, 2001.

\bibitem{giedke2001separability} Giedke, G., Kraus, B., Lewenstein, M. and Cirac, J. I., \emph{Separability properties of three-mode Gaussian states}, Physical Review A, {\bf 64}, 5, 2001.


\bibitem{braunstein2005quantum} Braunstein, S. L. and Van Loock, P., ``Quantum information with continuous variables'', \emph{Reviews of Modern Physics}, {\bf 77}, 2, 2005.

\bibitem{simon1988gaussian} Simon, R. Sudarshan, E. C. G. and Mukunda, N., ``Gaussian pure states in quantum mechanics and the symplectic group'', \emph{Physical Review A}, {\bf 37}, 8, 1988.


\bibitem{walls2007quantum} Walls, D. F. and Milburn, G. J., ``Quantum optics'', 2007, \emph{Springer Science \& Business Media}.

\bibitem{adesso2014continuous}   Adesso, G., Ragy, S. and Lee, A. R., ``Continuous variable quantum information: Gaussian states and beyond'', \emph{Open Systems \& Information Dynamics}, {\bf 21}, 2014, World Scientific.

\bibitem{ma1990multimode} Ma, X. and Rhodes, W., ``Multimode squeeze operators and squeezed states'', \emph{Physical Review A}, {\bf 41}, 9, 1990.

\bibitem{schnabel2017squeezed} Schnabel, R., ``Squeezed states of light and their applications in laser interferometers'', \emph{Physics Reports}, {\bf 684}, 2017, Elsevier.

\bibitem{pirandola2009correlation} Pirandola, S., Serafini, A. and Lloyd, S., ``Correlation matrices of two-mode bosonic systems'', \emph{Physical Review A}, {\bf 79}, 2009.

\bibitem{weedbrook2012gaussian} Weedbrook, C. et. al., ``Gaussian quantum information'', \emph{Reviews of Modern Physics}, {\bf 84}, 2, 2012.


\bibitem{Arvind:1995ab}  Arvind, B.~Dutta, N.~Mukunda and R.~Simon, ``The Real symplectic groups in quantum mechanics and optics,'' Pramana {\bf 45}, 471 (1995).

\bibitem{moshinsky1971linear}  Moshinsky, M. and Quesne, C., ``Linear canonical transformations and their unitary representations'', \emph{Journal of Mathematical Physics}, {\bf 12}, 8, 1971.

\bibitem{torre2005linear}  Torre, A., ``Linear ray and wave optics in phase space: bridging ray and wave optics via the Wigner phase-space picture'', 2005, Elsevier.

\bibitem{wolf2016development} Wolf, K., ``Development of linear canonical transforms: a historical sketch'', in \emph{Linear Canonical Transforms}, 3-28, 2016, Springer.

\bibitem{de2011symplectic} De Gosson, M. A. \emph{Symplectic methods in harmonic analysis and in mathematical physics}, {\bf 7}, 2011, {Springer Science \& Business Media}.


\bibitem{hariharan2005geometric} P. Hariharan, ``The geometric phase'', \emph{Progress in Optics}, {\bf 48}, 2005.

\bibitem{chruscinski2012geometric} D. Chruscinski and A. Jamiolkowski, \emph{Geometric phases in classical and quantum mechanics}, {\bf 36}, 2012, Springer Science \& Business Media.

\bibitem{sachdev2007quantum} S. Sachdev, \it{Quantum phase transitions}, \emph{Handbook of Magnetism and Advanced Magnetic Materials}, 2007, Wiley Online Library.

\bibitem{cabra2012modern} Daniel C. Cabra,  Andreas Honecker,  and P. Pujol, \emph{Modern theories of many-particle systems in condensed matter physics}, {\bf 843}, 2012, Springer Science \& Business Media.


\bibitem{chaturvedi1987berry} Chaturvedi, S., Sriram, M. S. and Srinivasan, V., ``Berry's phase for coherent states'', \emph{Journal of Physics A: Mathematical and General}, {\bf 20}, L1071, 1987.

\bibitem{Chiao} R. Y. Chiao and T. F. Jordan, ``Lorentz-group Berry phases in squeezed light'', \emph{Physics Letters A}, {\bf 132}, 77, 1988.

\bibitem{simon1993bargmann} R. Simon and N. Mukunda, ``Bargmann invariant and the geometry of the G\"uoy effect'', \emph{Physical review letters}, {\bf 70}, 7, 1993.

\bibitem{kwiat1991observation} P. G. Kwiat and R. Y. Chiao, ``Observation of a nonclassical Berry?s phase for the photon'', \emph{Physical review letters}, {\bf 66}, 5, 1991.

\bibitem{mukunda1993quantum} Mukunda, N. and Simon, R., ``Quantum kinematic approach to the geometric phase. I. General formalism'', \emph{Annals of Physics}, {\bf 228}, 2, 1993.

\bibitem{brendel1995geometric} J. Brendel, W. Dultz and W. Martienssen, ``Geometric phases in two-photon interference experiments'', \emph{Physical Review A}, {\bf 52}, 4, 1995.

\bibitem{strekalov1997two}  D. V. Strekalov and Y. H. Shih, ``Two-photon geometrical phase'', \emph{Physical Review A}, {\bf 56}, 4, 1997.

\bibitem{seshadri1997geometric} S. Seshadri, S. Lakshmibala and V. Balakrishnan, ``Geometric phases for generalized squeezed coherent states'', \emph{Physical Review A}, {\bf 55}, 2, 1997.

\bibitem{mendas1997pancharatnam} I. Mendas, ``Pancharatnam phase for ordinary and generalized squeezed states'', \emph{Physical Review A}, {\bf 55}, 2, 1997.

\bibitem{fuentes2000proposal} I. Fuentes-Guridi, S. Bose, and V. Vedral, ``Proposal for measurement of harmonic oscillator Berry phase in ion traps'', \emph{Physical review letters}, {\bf 85}, 24, 2000.

\bibitem{de2002squeezing} A. S. M. de Castro and V. V. Dodonov ``Squeezing exchange and entanglement between resonantly coupled modes'', \emph{Journal of Russian Laser Research},{\bf 23}, 2, 2002.










\bibitem{hall2018theory} Hall, M., ``The theory of groups'', 2018, \emph{Courier Dover Publications}

\bibitem{mukunda2014classical} N. Mukunda, S. Chaturvedi and R. Simon, ``Classical light beams and geometric phases'', \emph{JOSA A}, {\bf 31}, 6, 2014, \it{Optical Society of America}.





\end{thebibliography}
\end{document}